# Fairer Shootouts in Soccer: The *m* - *n* Rule

Steven J. Brams
Department of Politics
New York University
New York, NY 10012
USA
steven.brams@nyu.edu

Mehmet S. Ismail
Department of Political Economy
King's College London
London WC2R 2LS
UK
mehmet.ismail@kcl.ac.uk

D. Marc Kilgour
Department of Mathematics
Wilfrid Laurier University
Waterloo, Ontario N2L 3C5
CANADA
mkilgour@wlu.ca

## Abstract

Winning the coin toss at the end of a tied soccer game gives a team the right to choose whether to kick either first or second on all five rounds of penalty kicks, when each team is allowed one kick per round. There is considerable evidence that the right to make this choice, which is usually to kick first, gives a team a significant advantage. To make the outcome of a tied game fairer, we suggest a rule that handicaps the team that kicks first (*A*), requiring it to succeed on one more penalty kick than the team that kicks second (*B*). We call this the *m* - *n* rule and, more specifically, propose (*m*, *n*) = (5, 4): For *A* to win, it must successfully kick 5 goals before the end of the round in which *B* kicks its 4$^{th}$; for *B* to win, it must succeed on 4 penalty kicks before *A* succeeds on 5. If both teams reach (5, 4) on the same round—when they both kick successfully at (4, 3)—then the game is decided by round-by-round "sudden death," whereby the winner is the first team to score in a subsequent round when the other team does not. We show that this rule is fair in tending to equalize the ability of each team to win a tied game in a penalty shootout. We also discuss a related rule that precludes the teams from reaching (5, 4) at the same time, obviating the need for sudden death and extra rounds.

## 1. **Introduction**

Soccer, the most widely played sport in the world, is known as football except in North America. In tournaments and elimination competitions that must end with a single winner, matches that end in a tie after regulation play (90 minutes) and one overtime period (30 minutes), the tie is broken by a penalty shootout, wherein the team that wins a coin toss has the right to choose whether to kick first or second on all five rounds of penalty kicks. Although it remains unclear whether kicking first or second is advantageous (evidence, cited shortly, is mixed), there is considerable evidence that the right to make this choice, which is usually to kick first, gives the team that makes it a significant advantage.

To make the outcome of a tied game fairer, we suggest a rule that handicaps the team that kicks first (*A*), requiring it to succeed on more penalty kicks than the team that kicks second (*B*). We call this the *m* - *n* rule and, more specifically, propose (*m*, *n*) = (5, 4): For *A* to win, it must successfully kick 5 goals before the end of the round in which *B* kicks its $4^{th}$; for *B* to win, it must succeed on 4 penalty kicks before *A* succeeds on 5. If the teams reach (5, 4) on the same round—when they both kick successfully at (4, 3)—then the game is decided by round-by-round "sudden death," whereby the winner is the first team to score in a subsequent round when the opponent does not. Table 1 uses three examples to illustrate what can happen in shootouts under the (5, 4) rule. We also discuss a related rule that precludes the teams from reaching (5, 4), obviating the need for sudden death and extra rounds.[1]

[1] In the subsequent analysis, we assume $m = n + 1$, but the analysis can readily be extended to $m > n + 1$ to model situations in which *B* may need a greater handicap to countervail *A*'s advantage in kicking first.

| | Round 1 | Round 2 | Round 3 | Round 4 | Round 5 | Result |
|---|---|---|---|---|---|---|
| Team *A* | ✓ | ✓ | ✓ | ✓ | ✓ | *A* wins 5-3 |
| Team *B* | ✓ | ✓ | ✓ | ✗ | ✗ | |
| Team *A* | ✓ | ✓ | ✓ | ✓ | ✗ | *B* wins 4-4 |
| Team *B* | ✓ | ✓ | ✓ | ✗ | ✓ | |
| Team *A* | ✓ | ✓ | ✓ | ✓ | ✓ | Sudden Death 5-4 |
| Team *B* | ✓ | ✓ | ✓ | ✗ | ✓ | |

**Table 1. Three shootout examples under the (5, 4) rule, wherein a checkmark indicates scoring and the letter *x* failing to score**

Is choosing to kick first beneficial? The evidence provided by Apesteguia and Palacios-Huerta (2010) from international soccer tournaments indicates that kicking first gave teams about a 60% win advantage, which is significant. However, later evidence of Kocher et al. (2012), using a larger sample of games, indicated that first-kicking teams had only a 53.3% win advantage, which was not significant. Arrondel et al. (2019) found that the first-kicking team won 50.4% of 252 shootouts in French competitions. In a subsequent study, Rudi et al. (2020) extended the previous research (excluding French cups) by analyzing a larger dataset of 1623 games; they found that the win advantage of the first-kicking team was 55%, which is statistically significant. This translates into a 22% greater probability of winning the shootout for the team that kicks first.

Still later, Kassis et al. (2021) argued that it is winning the coin toss, and thereby being able to choose whether to kick first or second, that confers the real advantage. Evidence from international tournaments between 2003 and 2017 indicates that the team that wins the coin toss, and so makes the choice of whether to kick first or second, is key: Teams that won the coin toss won 66% of 65 shootouts, whether they kicked first or second.

Kassis et al. (2021) hypothesized that if a team's goalie is not as good as one's opponent's, it is better to kick first to prevent one's opponent from jumping ahead early; but if a team's goalie is better than the opponent's, it is better to kick second to try to catch up and go ahead. As the authors put it, "being 'behind schedule' may be the decisive aspect that generates psychological pressure" (p. 281), which causes the behind team to choke and leads to a loss.

At the most competitive level of soccer—the FIFA World Cup—the pressure is greatest on a team when an opponent takes the lead toward the end of a shootout. Kicking first, on balance, seems the better way to create such pressure. In fact, in 10 of the last 11 shootouts in FIFA World Cups, teams that won the coin toss—including the national team of Argentina that won the shootout in the championship game in the 2022 FIFA World Cup in Qatar—chose to kick first (9 of these recent shootouts were not included in the dataset of Kassis et al., 2021). At least in these high-stakes tournaments, the winner of the coin toss almost always chooses to kick first.

In a survey of 340 professional and amateur coaches from Germany, Switzerland, and Austria, Kassis et al. (2021) found that an overwhelming majority—88% of respondents—recommended that their captain should choose to kick first if they won the coin toss. This finding is consistent with the previous survey by Apesteguia and Palacios-Huerta (2010), who found that more than 90% of coaches and players recommended kicking first.

Accordingly, we assume in the subsequent analysis that if the winner of the coin toss chooses to kick first (assume it is *A*), it puts itself in an advantageous position. Therefore, to try to equalize the two teams' positions, we propose the (5, 4) rule under which *B* needs to kick one fewer goal to win the shootout.

Occasionally, the winner of a coin toss chooses to kick second and so plays the role of *B*. Not having to succeed as often, according to the (5, 4) rule, will help to balance *A*'s usual advantage in kicking first and make the option of kicking second more desirable.

Should the (5, 4) rule be implemented, we do not expect the lopsided choices made by teams in FIFA World Cup shootouts—overwhelmingly favoring *A*—to occur. Instead, we conjecture that the winners of the coin toss will split more evenly between choosing to be *A* (kicking first) or *B* (kicking second).

To summarize so far, if *A* kicks first, it wins if at the end of a round it has scored $m$ goals and *B* has scored fewer than $n$ goals, whereas *B* wins if at the end of a round it has scored $n$ goals and *A* has scored fewer than $m$. If the score is exactly $(m, n)$ at the end of a round, the shootout goes to sudden death. On a sudden-death round, if one team scores and the other team does not, then the scoring team wins. If both teams score or neither team does, then there is another sudden-death round, and so on ad infinitum until one team scores and the other does not.

## 2. Comparison with Regular Soccer Shootout

In regular soccer shootouts, a five-round shootout may end early if one team is so far ahead that the behind team cannot win, even if it succeeds on all its remaining kicks while the ahead team fails. For example, if the score is (4, 2) for (*A*, *B*) after four rounds, there is no way for *B* to catch up to *A*, much less win, even if *B* scores and *A* doesn't on the fifth round, bringing the shootout to (4, 3). Thus, at (4, 2) on the fourth round, one can declare *A* the winner and skip the fifth round.

This is not true under the $(m, n) = (5, 4)$ rule. At (4, 2), *A* has not yet scored the 5 goals it needs to win (before *B* obtains 4 goals), and *B* has not yet scored the 4 goals it needs to win (before *A* obtains 5 goals). This is also true if the score goes from (4, 2) to (4, 3). But if a round ends at (5, 4), *A* and *B* must go to a tiebreaker, as noted earlier, because even though both teams have reached their required number of goals to win, they did so on the same round.

The $(m, n)$ rule reduces the role of luck in a shootout, whereby one team wins with a string of successes while the other team fails. For *A*, the earliest it can win under the (5, 4) rule is at (5, 0), and for *B* it is at (0, 4). We estimate that the shootout would last on average about six rounds (for more discussion, see section 3). By comparison, in regular soccer, a shootout typically lasts five rounds, though a team can win a shootout in as few as 3 rounds if it succeeds on all its kicks while the other team fails. These examples illustrate that there cannot be as early a wipeout of one team in a (5, 4) shootout as there can be in regular soccer.

To illustrate the intuition behind the $(m, n)$ rule, we begin by showing how it works in the simple case of $(m, n) = (2, 1)$. We represent this shootout as the 10-state absorbing Markov chain shown in Figure 1. Five of the states are labelled *X:* $(m, n)$, where $X = A$ or $B$ is the team to shoot next, and $(m, n)$ is the current score. Each shot results in a move from one state to another, indicated by an arrow. Notice that the two "win" states in Figure 1, labelled "*A* wins" and "*B* wins," are *absorbing;* there are no outgoing arcs.

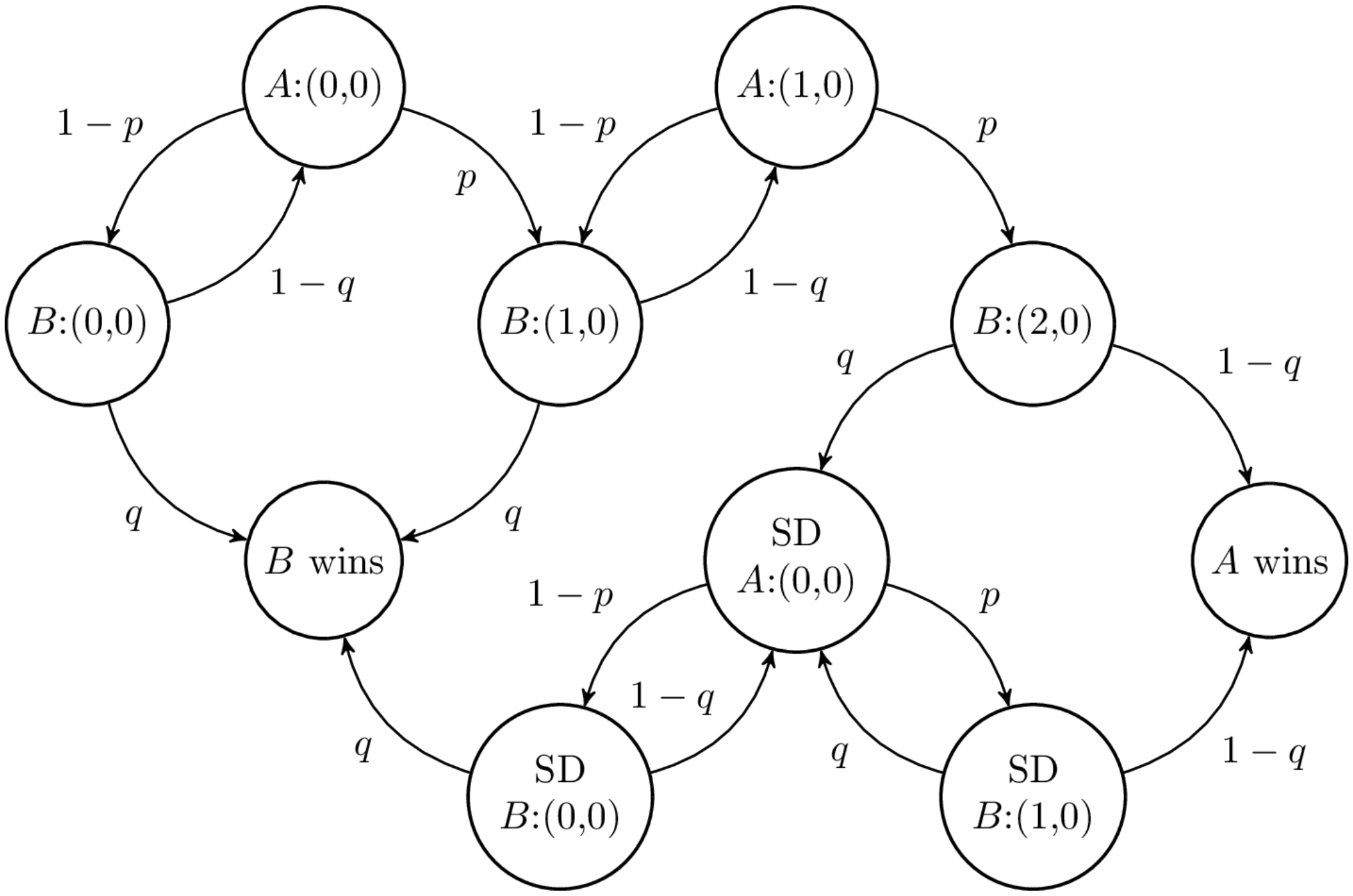

**Figure 1. The $(2,1)$ shootout Markov chain**

In this model, $A$ kicks first, so the initial state is $A$:(0, 0). To win, $A$ must score twice before $B$ scores. For $B$ to win, it must score once before $A$ scores twice. We assume that kicks are independent events; a kick by $A$ is successful with probability $p$, and a kick by $B$ is successful with probability $q$. Kicks occur in rounds. If after a round the score is (2, 1), then the contest is decided by sudden death; it enters the subchain consisting of the three SD states, and does not leave this subchain until it is absorbed.

In fact, a Sudden Death shootout has already been analyzed. The probability that $A$ wins is $\frac{p(1-q)}{p+q-2pq}$ (see Brams and Ismail, 2018, p. 192); of course, $B$ wins with the complementary probability. For any other non-absorbing state, $S$, we define the probability that $A$ wins, conditional on just having entered state $S$, to be

$$P(S) = Pr\ (A \text{ wins} \mid S).$$

If one starts at $S = B$: (1, 0), $B$ scores its first goal with probability $q$—and so wins the game—or does not score with probability $(1 - q)$, in which case the state becomes $A$: (1, 0). It follows that

$$P(B{:}\,1,0) = q\ (0) + (1 - q)\ P(A{:}\,1,0) \tag{1}$$

Now start at $A$: (1, 0). If $A$ scores its second goal with probability $p$, and $B$ does not score its first goal with probability $(1 - q)$, $A$ wins. If both $A$ and $B$ score, which has probability $pq$, the chain proceeds to a shootout. If $A$ does not score (probability $1 - p$), the state becomes $B$: (1, 0). Thus, $A$'s win probability from $A$: (1, 0) is given by the following formula:

$$P(A{:}\,1,0) = p(1-q)(1) + pq\left(\frac{p(1-q)}{p+q-2pq}\right) + (1 - p)P(B{:}\,1,0) \tag{2}$$

Solving equations (1) and (2) simultaneously yields

$$P(A{:}\,1,0) = \frac{p(1-q)}{p+q-2pq};\ P(B{:}\,1,0) = \frac{p(1-q)^2}{p+q-2pq}.$$

At state $A{:}\,(0,0)$, it follows that

$$\begin{aligned} P(A{:}\,0,0) &= p\ P(B{:}\,1,0) + (1 - p)P(B{:}\,0,0) \\ &= p\ P(B{:}\,1,0) + (1 - p)(1 - q)\ P(A{:}\,0,0). \end{aligned} \tag{3}$$

Substituting (1) into (3) and solving for $P(A{:}0, 0)$ yields[2]

$$P(A{:}\,0,0) = \frac{p^2(1-q)^2}{(p+q-2pq)(p+q-pq)}.$$

---

[2] Standard approaches for analyzing absorbing Markov chains using the transition matrix yield the same result.

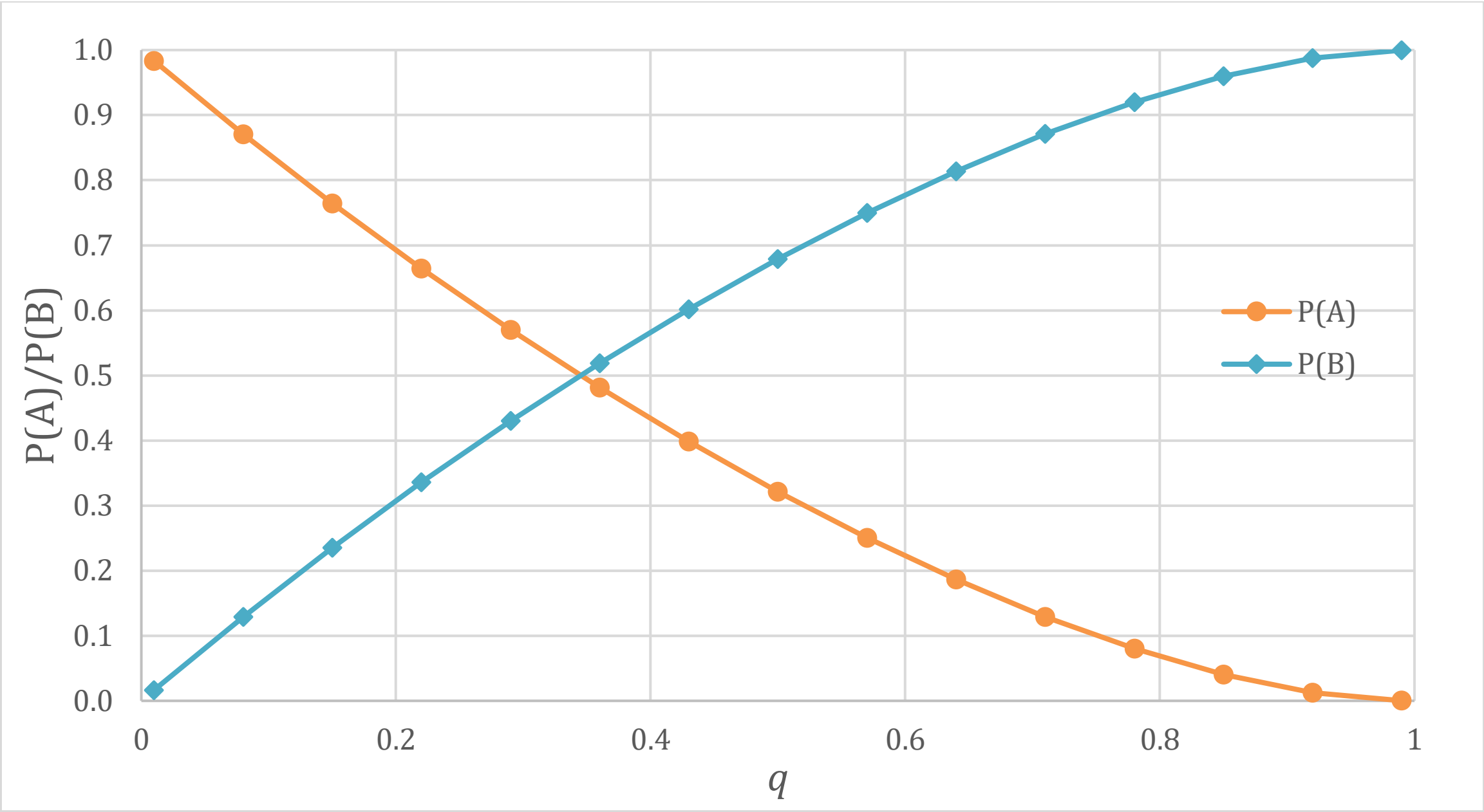

**Figure 2. (2, 1) rule: $P(A)$ and $P(B)$ as a function of $q$ when $p = 0.75$**

Let $P(A) = P(A{:}\,0, 0)$ denote $A$'s win probability so that $P(B) = 1 - P(A)$ denotes $B$'s win probability. Figure 2 illustrates $P(A)$ and $P(B)$ as functions of $q$ when $p = 0.75$, a value that approximates the average success rate of kickers in professional soccer (Kassis, 2021, p. 279). Clearly, as $q$ approaches 1, $P$(A) decreases and $P(B)$ increases; at $q = 0.34$, $P(A) = P(B)$, which is to say that $B$ need only achieve about half the success rate of $A$ to give the two teams an equal probability of winning the shootout when $(m, n) = (2, 1)$. This, we believe, gives $B$ an undeserved advantage—far greater than it needs—to offset $A$'s advantage in winning the coin toss and, almost always, electing to kick first.

We next calculate the expected number of rounds the shootout lasts. Does sudden death, after a tie of (2, 1), inordinately lengthen the shootout? Let $ER(S)$ denote the expected number of rounds remaining in the shootout conditional on it having just entered state $S$ of the Markov chain shown in Figure 1:

$$ER(S) = \mathbb{E}(\text{\# rounds the shootout lasts} \mid S).$$

First, we calculate, $ER(A{:}\,1,0)$, which is the expected number of rounds of the shootout when $A$ is about to shoot and the current score is (1, 0):

$ER(A{:}\,1,0)$

$= [p(1-q) + (1-p)q](1) + (1-p)(1-q)(ER(A{:}\,1,0) + 1) + pq(ER(SD) + 1).$

The first summand, $[p(1-q) + (1-p)q](1)$, indicates that when one team scores and the other misses and, the shootout ends after one round. The second summand, $(1-p)(1-q)(ER(A{:}\,1,0) + 1)$, indicates that, when both $A$ and $B$ miss, the number of rounds in the shootout increases by 1. The last summand, $pq(ER(SD) + 1)$, indicates that when both $A$ and $B$ score, the shootout goes to sudden death after the current round. To work out $ER(SD)$, recall that, having entered the SD states, the chain remains there until absorbed. Therefore

$$ER(SD) = [p(1-q) + (1-p)q](1) + [pq + (1-p)(1-q)](ER(SD) + 1)$$

which produces

$$ER(SD) = \frac{1}{p + q - 2pq}.$$

Now solve the equation above for $ER(A{:}\,1,0)$ to obtain

$$ER(A{:}\,1,0) = \frac{1}{p + q - 2pq}.$$

Finally,

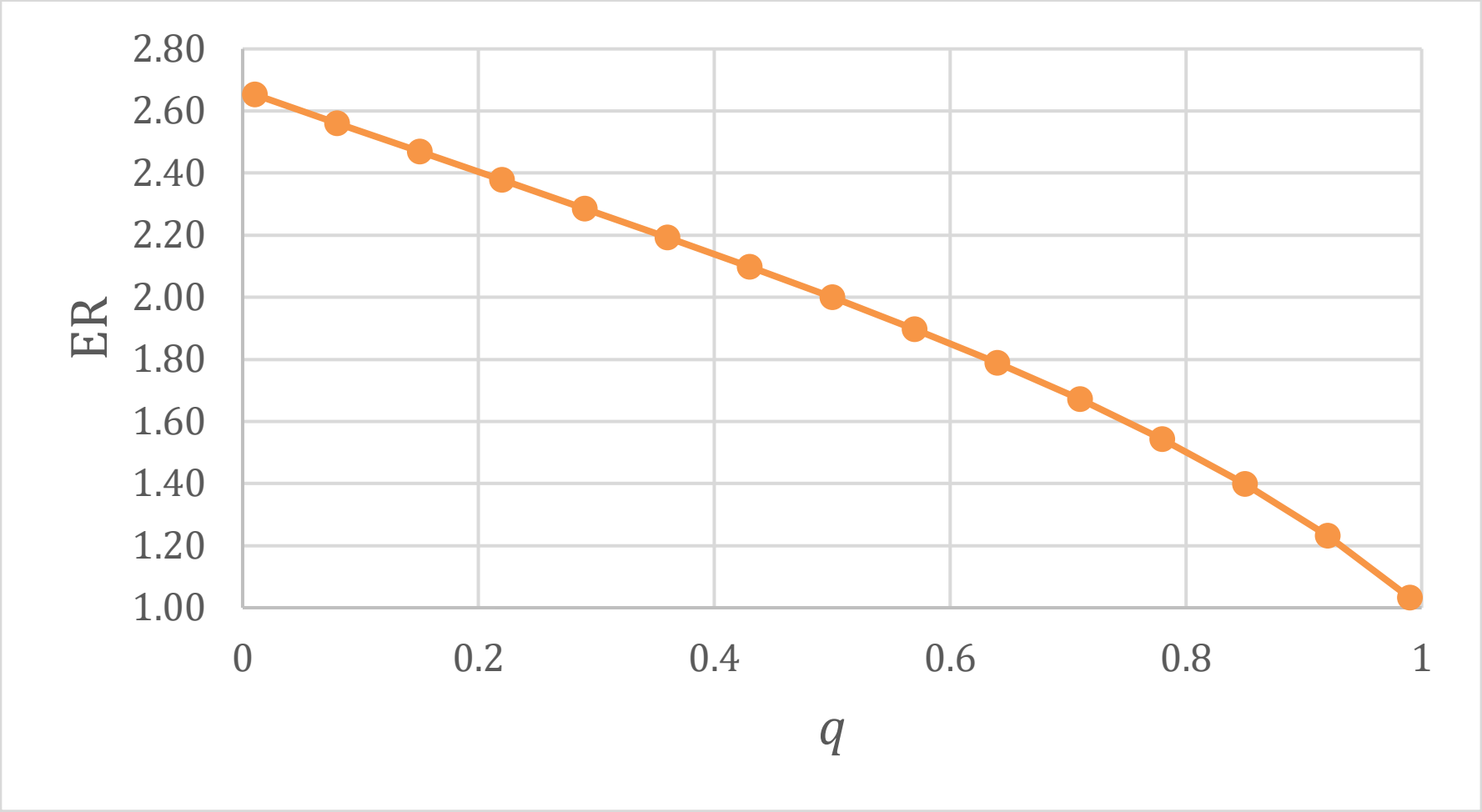

**Figure 3.** ***ER*** **as a function of** ***q*** **for** $p = 0.75$

$ER(A{:}\,0,0)$

$= p(1-q)(ER(A{:}\,1,0)+1) + [pq + (1-p)q](1)$

$+ (1-p)(1-q)(ER(A{:}\,0,0)+1)$

$= p(1-q)\left(1 + \frac{1}{p+q-2pq}\right) + q + (1-p)(1-q)(ER(A{:}\,0,0)+1).$

The first summand, $p(1-q)\left(1+\frac{1}{p+q-2pq}\right)$, indicates that when $A$ scores and $B$ misses, the expected rounds is given by $1 + ER(A{:}\,1,0)$. The second summand, $pq$, indicates that when both $A$ and $B$ score and, $B$ wins the contest in one round. The third summand, $(1-p)q$, indicates that when $A$ misses and $B$ scores, $B$ wins in one round. The last summand, $(1-p)(1-q)(ER(A{:}\,1,0)+1)$, indicates that when both $A$ and $B$ miss, the number of rounds increases by 1. Solving for $ER(A{:}\,0,0)$ yields

$$ER(A{:}\,0,0) = \frac{2p+q-3pq}{(p+q-2pq)(p+q-pq)}.$$

Figure 3 illustrates $ER = ER(A{:}\,0,0)$ under the $(2,1)$ rule for a fixed $p = 0.75$ as a function of $q$.

Note that as *B*'s probability of a successful kick approaches 1, *ER* approaches 1, despite the fact that *A* kicks first. This is because *B*'s successful kick on the first round ends the game, rendering *A*'s first kick irrelevant.

When $p$ = 0.75, Figure 3 shows that if *B* duplicates *A*'s probability of a successful kick ($q$ = 0.75), the match will, on average, take 1.60 rounds. If *A* kicks successfully on the first round and *B* does not, then *A* will have an opportunity to win the match at 2-0 if *B* fails again. But if both *A* and *B* succeed on the second round (with probability (3/4)(3/4) = 9/16 = 0.5625), the score will be 2-1, so the shootout will go to sudden death.

We next provide a general analysis of the ($m$, $n$) rule that extends the foregoing (2, 1) analysis. It shows that when ($m$, $n$) = (5, 4), the $q$ that equalizes the win probabilities of *A* and *B* is 0.60 when $p$ = 0.75, making the shootout fair when *B*'s win probability is 80% of *A*'s.

## 3. General Analysis of the ($m$, $n$) Rule

Let $\mathbb{N} = \{1, 2, \ldots\}$. We assume that $m \in \mathbb{N}$, $n \in \mathbb{N}$, $m > n$, $1 > p > 0$, and $1 > q > 0$. Consider round $r \in \mathbb{N}$ and recall that *A* kicks first and *B* kicks second. We assume that each kick is an independent event. Let $P_A(m, n, p, q)$ denote *A*'s win probability as a function of $m, n, p$, and $q$. Table  summarizes the notation used throughout the paper.

| *Notation* | *Meaning* |
|---|---|
| $A$ | First kicker in a round |
| $B$ | Second kicker in a round |
| $m = 1, 2, \ldots$ | Number of goals $A$ needs to win |
| $n = 1, 2, \ldots$ | Number of goals $B$ needs to win |
| $p \in (0,1)$ | Probability $A$ scores |
| $q \in (0,1)$ | Probability $B$ scores |
| $r = 1, 2, \ldots$ | Number of the current round |
| $P_X(m, n, p, q)$ | Win probability of $X = A$ or $B$ |
| $ER(m, n, p, q)$ | Expected number of rounds |

**Table 2. Notation used in calculations**

## The Probability of Winning

We next present a formula for $P_A(m, n, p, q)$.

**Proposition 1**. *A's win probability under the* $(m, n)$ *rule is given by the following expression.*

$$P_A(m, n, p, q)$$

$$= \sum_{r=m}^{\infty} \left[ \binom{r-1}{m-1} \, p^m (1-p)^{r-m} \left( \sum_{i=0}^{n-1} \binom{r}{i} \, q^i \, (1-q)^{r-i} + \binom{r-1}{n-1} \, q^n \, (1-q)^{r-n} \frac{p(1-q)}{p+q-2pq} \right) \right].$$

**Proof**. For each $r \in \{m, m+1, m+2, \ldots\}$, let $E_r^A$ denote the event that $A$ wins in exactly $r$ rounds (with no sudden death) and $SD_r^A$ denote the event that $A$ wins a sudden death that begins at the end of round $r$ with the shootout tied at $(m, \ n)$. Define $E_r := \ E_r^A \cup SD_r^A$, and note that $P(E_r) = P(E_r^A) + (SD_r^A)$ because $E_r^A$ and $SD_r^A$ are mutually exclusive events. Note also that

$$P_A(m,n,p,q) = P\left(\bigcup_{r\geq m} E_r\right) = \sum_{r\geq m} P(E_r),$$

because $E_m, E_{m+1}, E_{m+2}$ ... are mutually exclusive events.

The proof strategy is to compute for every $r$ the probability of $E_r$ and then sum these probabilities. We next derive the desired formula for $P_A(m,n,p,q)$ in five main steps.

**Step 1**: The probability that *A* scores $m$ goals in exactly $r$ rounds is given by

$$\binom{r-1}{m-1} p^m (1-p)^{r-m}.$$

This is because *A* must score $m-1$ goals in $r-1$ rounds plus a goal on round $r$. Thus, there are $\binom{r-1}{m-1}$ ways in which *A* scores $m$ goals in $r$ rounds. Because *A* scores $m$ goals in total and each of these events are independent, we have the factor $p^m$. In the remaining $r-m$ rounds, *A* does not score; hence, the factor $(1-p)^{r-m}$.

**Step 2**: The probability that *B* scores $n-1$ or fewer goals in $r$ rounds is given by

$$\sum_{i=0}^{n-1} \binom{r}{i} q^i (1-q)^{r-i}.$$

**Step 3**: Given that *A* scores *m* goals in *r* rounds, the conditional probability that there is a sudden death segment is the probability that *B* has scored exactly *n* goals as of the end of round $r$,

$$\binom{r-1}{n-1} q^n (1-q)^{r-n}.$$

**Step 4**: As observed above, the probability that *A* wins in sudden death is

$$\frac{p(1-q)}{p+q-2pq}.$$

**Step 5**: Combining Steps 1-4, $A$ wins if and only if $A$ scores $m$ goals in $r$ rounds, for some $r \geq m$, and either $B$ scores fewer than $n$ goals in $r$ rounds or $B$ scores exactly $n$ goals in $r$ rounds and $A$ wins the sudden death segment. The index of the summation in the formula for $P_A(m,n,p,q)$ starts at $r = m$, because $A$ can win the shootout in $m$ or more rounds but cannot win it in fewer than $m$ rounds. Q.E.D.

The probability that $B$ wins, $P_B(m,n,p,q)$, can be obtained by carrying out a procedure similar to the proof of Proposition 1, or by reversing the roles of the two players in the formula for $P_A(m,n,p,q)$.

**Corollary 1**. *B's win probability under the* $(m,n)$ *rule is given by the following expression.*

$$P_B(m,n,p,q)$$

$$= \sum_{r=n}^{m-1} \left[ \binom{r-1}{n-1} \; q^n (1-q)^{r-n} \right]$$

$$+ \sum_{r=m}^{\infty} \left[ \binom{r-1}{n-1} \; q^n (1-q)^{r-n} \left( \sum_{i=0}^{m-1} \binom{r}{i} \; p^i \, (1-p)^{r-i} \right.\right.$$

$$\left.\left. + \binom{r-1}{m-1} \; p^m \, (1-p)^{r-m} \frac{q(1-p)}{p+q-2pq} \right) \right].$$

This corollary follows from interchanging $p$ with $q$ and $m$ with $n$ and splitting the summation (because $m > n$) in the formula for $P_A(m,n,p,q)$.

**Remark 1.** Note that $A$'s win probability and $B$'s win probability must sum to 1 because the contest cannot end in a tie. Hence, we have the following equality:

$$P_A(m,n,p,q) + P_B(m,n,p,q) = 1.$$

| $m$ | $n$ | $p$ | $q^*(p)$ |
|---|---|---|---|
| 5 | 4 | 0.75 | 0.60 |
| 4 | 3 | 0.75 | 0.56 |
| 3 | 2 | 0.75 | 0.50 |
| 2 | 1 | 0.75 | 0.34 |

**Table 3. *B*'s balancing probability *q* when $p = 0.75$**

Fix $m$ and $n$. For any $p$, define $q^*(p)$ to be the value of $q$ that makes $P_A(m, n, p, q) = P_B(m, n, p, q)$. We call $q^*(p)$ the *balancing probability* for $B$, as it renders $P_B = P_A$. Table gives, for different values of $m$ and $n$, the values of $q^*(0.75)$ , i.e., the value of $q$ that equalizes the two win probabilities. When $(m, n) = (5, 4)$, observe that $q^*(0.75) = 0.60$, 80% of the value of $p = 0.75$ (for $A$) and, we think, a reasonable approximation of the lower success probability of kicking second.[3]

Figure 4 illustrates the graphs of $P_A$ and $P_B$ for $p = 0.75$ and all possible values of $q$. Notice that these curves intersect at $q = 0.60$, the balancing probability for $B$ when $p = 0.75$. In particular, when $(m, n) = (5, 4)$, $A$'s advantage in kicking first ($p = 0.75$ vs. $q = 0.60$ for $B$) is exactly balanced by $B$'s advantage in having to score one less goal to win the shootout.

[3] Recall that the first-kicking team has about 22% higher probability of winning the shootout than the second-kicking team (Rudi et al., 2021).

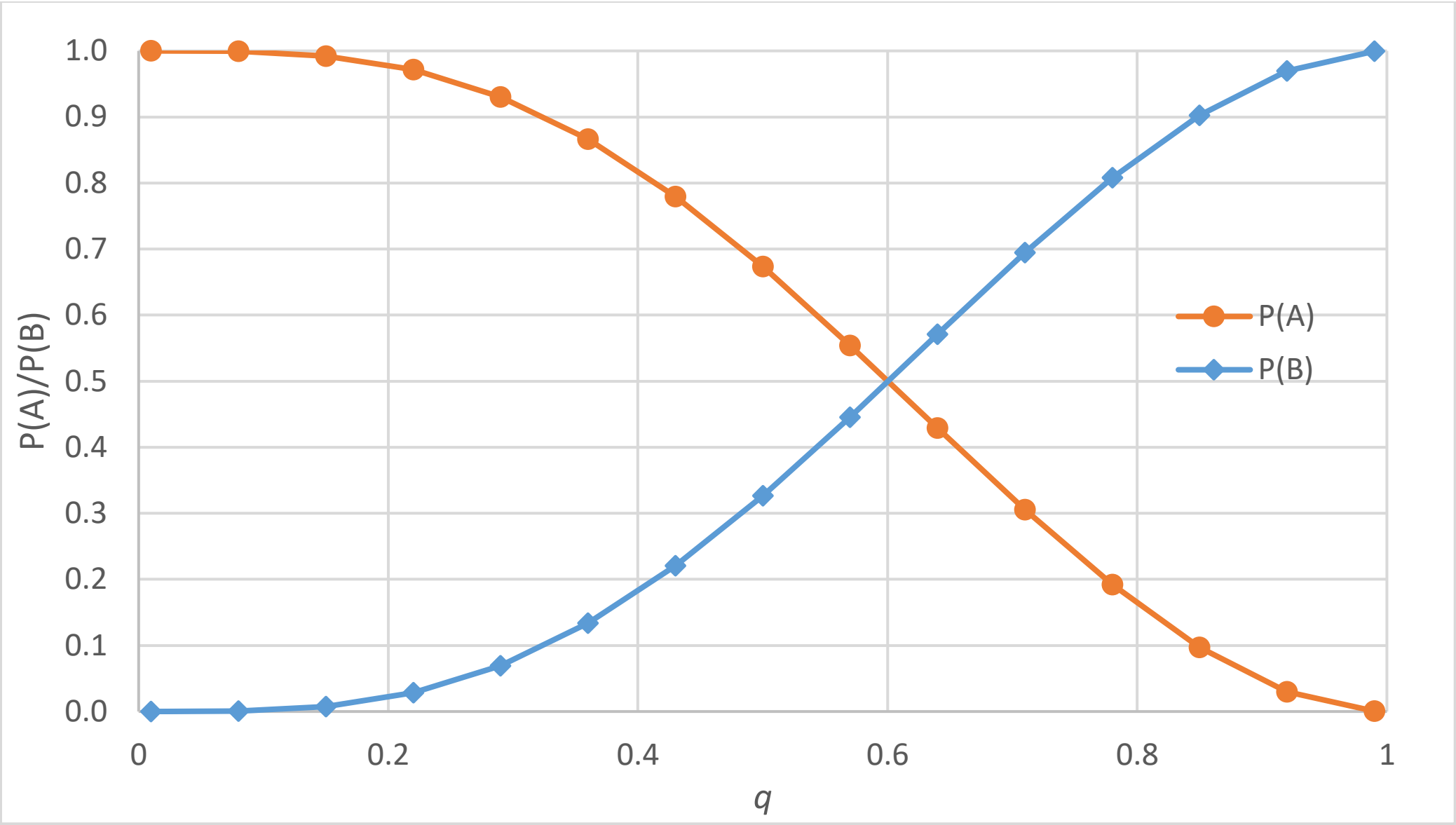

**Figure 4. P(A):= $P_A(5, 4, 0.75, q)$ and P(B):= $P_B(5, 4, 0.75, q)$ as a function of $q$**

**The Expected Number of Rounds**

Let $ER(m, n, p, q)$ denote the expected number of rounds of the shootout under the $(m, n)$ rule.

**Proposition 2**. *The formula for* $ER(m, n, p, q)$ *is given by the following expression.*

$$ER(m, n, p, q)$$

$$= \sum_{r=m}^{\infty}\left[r\binom{r-1}{m-1} p^m(1-p)^{r-m}\left(\sum_{i=0}^{n-1}\binom{r}{i} q^i (1-q)^{r-i}\right)\right]$$

$$+ \sum_{r=n}^{\infty}\left[r\binom{r-1}{n-1} q^n(1-q)^{r-n}\left(\sum_{i=0}^{\max\{r,\ m-1\}}\binom{r}{i} p^i (1-p)^{r-i}\right)\right]$$

$$+ \sum_{r=m}^{\infty}\left[\left(r+\frac{1}{p+q-2pq}\right)\binom{r-1}{m-1} p^m(1-p)^{r-m}\binom{r-1}{n-1} q^n (1-q)^{r-n}\right].$$

**Proof**. For each $r \in \{n, n+1, n+2, \ldots\}$, let $E_r^B$ denote the event that $B$ wins in exactly $r$ rounds and $SD_r$ denote the event that the shootout goes to sudden death at the end of round $r$. Recall that $E_r^A$ denotes the event that $A$ wins in exactly $r$ rounds. The proof strategy is to find the expected number of rounds by multiplying the number of rounds of each event by the probability of that event and summing these expectations over the rounds:

$$ER(m,n,p,q) = \sum_{r \geq m} rP(E_r^A) + \sum_{r \geq n} rP(E_r^B) + \sum_{r \geq m} \big(r + ER(SD)\big)P(SD_r).$$

We derive the formula for $ER(m,n,p,q)$ in three steps.

**Step 1**: Sum over $r \geq m$ the product of $r$ and the probability of $A$ winning in $r$ rounds with no shootout, as calculated in the proof of Proposition 1:

$$\sum_{r=m}^{\infty} \left[ r \binom{r-1}{m-1} p^m (1-p)^{r-m} \left( \sum_{i=0}^{n-1} \binom{r}{i} q^i (1-q)^{r-i} \right) \right].$$

**Step 2**: Sum over $r \geq n$ the product of $r$ and the probability of $B$ winning in $r$ rounds with no shootout,

$$\sum_{r=n}^{\infty} \left[ r \binom{r-1}{n-1} q^n (1-q)^{r-n} \left( \sum_{i=0}^{\max\{r,\ m-1\}} \binom{r}{i} p^i (1-p)^{r-i} \right) \right].$$

**Step 3**: Sum over $r \geq m$ the product of $\left(r + \frac{1}{p+q-2pq}\right)$ and the probability that the contest goes to sudden death beginning after round $r$, in other words that the score is $(m,n)$ after $r$ rounds. The expected number of rounds of the shootout in the sudden death segment is given by

$$\frac{1}{p + q - 2pq},$$

which is calculated in section 2. Therefore, we obtain the following summation in this step

$$\sum_{r=m}^{\infty} \left[ \left( r + \frac{1}{p + q - 2pq} \right) \binom{r-1}{m-1} \; p^m (1-p)^{r-m} \binom{r-1}{n-1} \; q^n \; (1-q)^{r-n} \right].$$

| $m$ | $n$ | $ER(m, n, 0.75, 0.6)$ |
|---|---|---|
| 5 | 4 | 6.06 |
| 4 | 3 | 4.71 |
| 3 | 2 | 3.33 |
| 2 | 1 | 1.85 |

**Table 4.** $\boldsymbol{ER}$ **for different values of** $\boldsymbol{m}$ **and** $\boldsymbol{n}$

As a result, we obtain the desired formula for $ER(m, n, p, q)$. Q.E.D.

Table 4 illustrates $ER$ for different values of $(m, n)$. Observe that when $(m, n)$ = (5, 4), the expected number of rounds of a shootout is slightly more than 6 rounds.

**Remark 2.** To double check the $ER$ formula, we can confirm that $ER(2,1,p,q)$ reduces to

$$\frac{2p + q - 3pq}{(p + q - 2pq)(p + q - pq)},$$

which we obtained in section 2. For $p$ = 0.75, this formula gave $ER = 1.60$ when $q$ = 0.75, but when $q$ = 0.60, which equalizes $A$'s and $B$'s win probabilities, $ER$ increases to 1.85 because of the greater probability of having to go to sudden death when $A$ and $B$ have equal win probabilities.

It is useful to compare the results we have obtained so far with the following modification of our model. Assume that the teams alternate kicking and the first team that reaches its target, $m$ or $n$, wins the shootout. This model is not based on rounds but, instead, on alternating kicks until either $A$ or $B$ becomes the first team to reach $m$ or $n$, respectively. Because both teams cannot reach their targets simultaneously when they do not kick in rounds, we refer to this as the *sequential model*.

Let $Q(A)$ and $ER(Q)$ denote $A$'s win probability and the expected number of rounds of the shootout under the sequential model. Figure 5 and Figure 6 illustrate that this model does not make a big difference in terms of win probability or expected number of rounds of the shootout. We prefer to keep rounds and, therefore, sudden death—in which either team can win should a tied game reach this phase in a penalty shootout—because it is more exciting to fans than letting one team, because it happens to kick next, be the only team that can "seal the deal." Nonetheless, the sequential shootout model offers an alternative way of conducting a shootout that is easy to implement and follow.

Both the standard model and the sequential model are *strategyproof* in the sense that neither team can increase its winning probability by deliberately missing a kick, because this decreases its probability of attaining the required number of goals to win without changing the order of kicking. For an example of a strategy-vulnerable sequential rule, see, e.g., Brams et al. (2018).

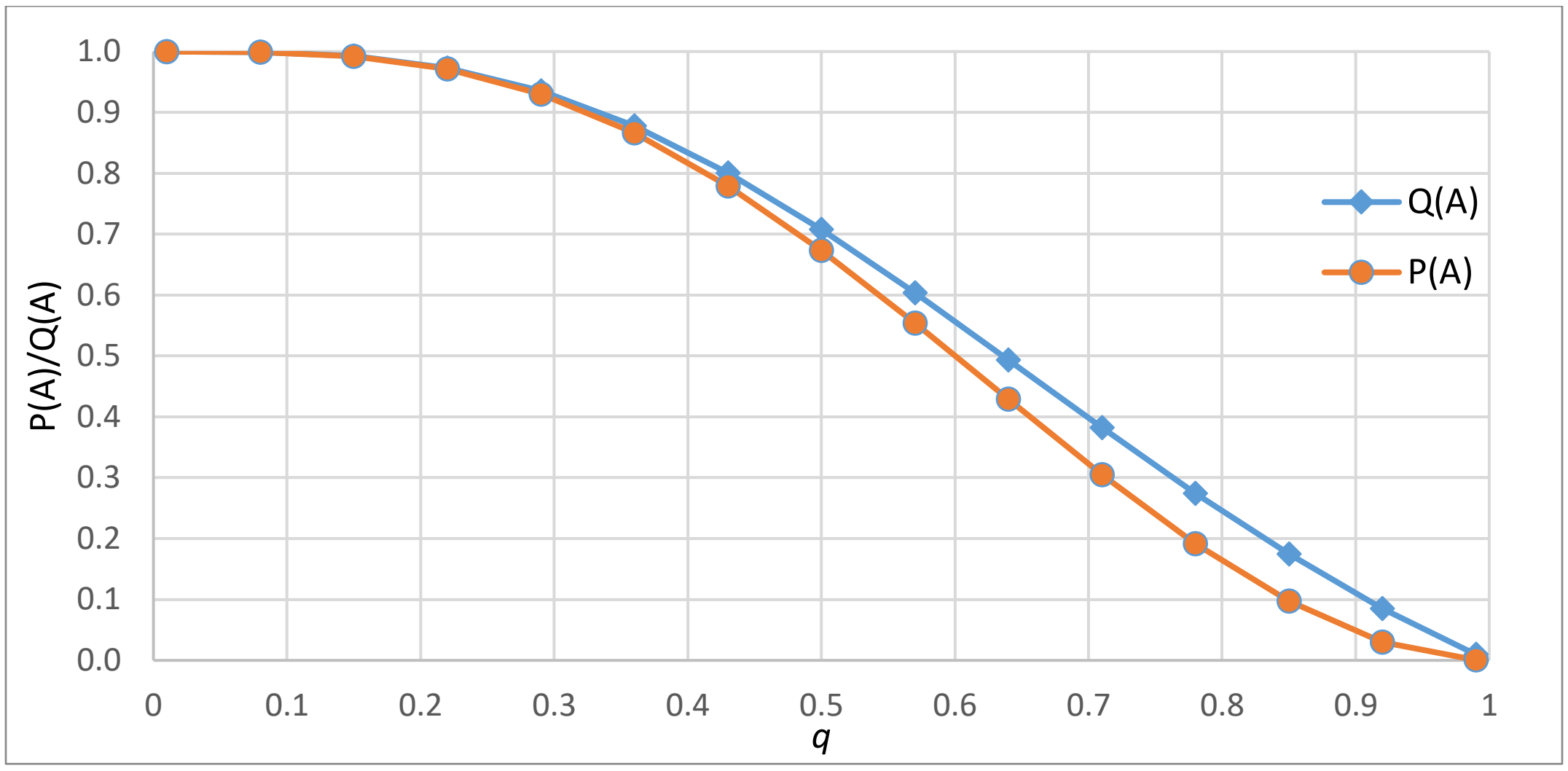

**Figure 5.** $\boldsymbol{P(A)}$**, and** $\boldsymbol{Q(A)}$ **for the sequential model, under the 5-4 rule as a function of** $\boldsymbol{q}$**, given** $\boldsymbol{p = 0.75}$

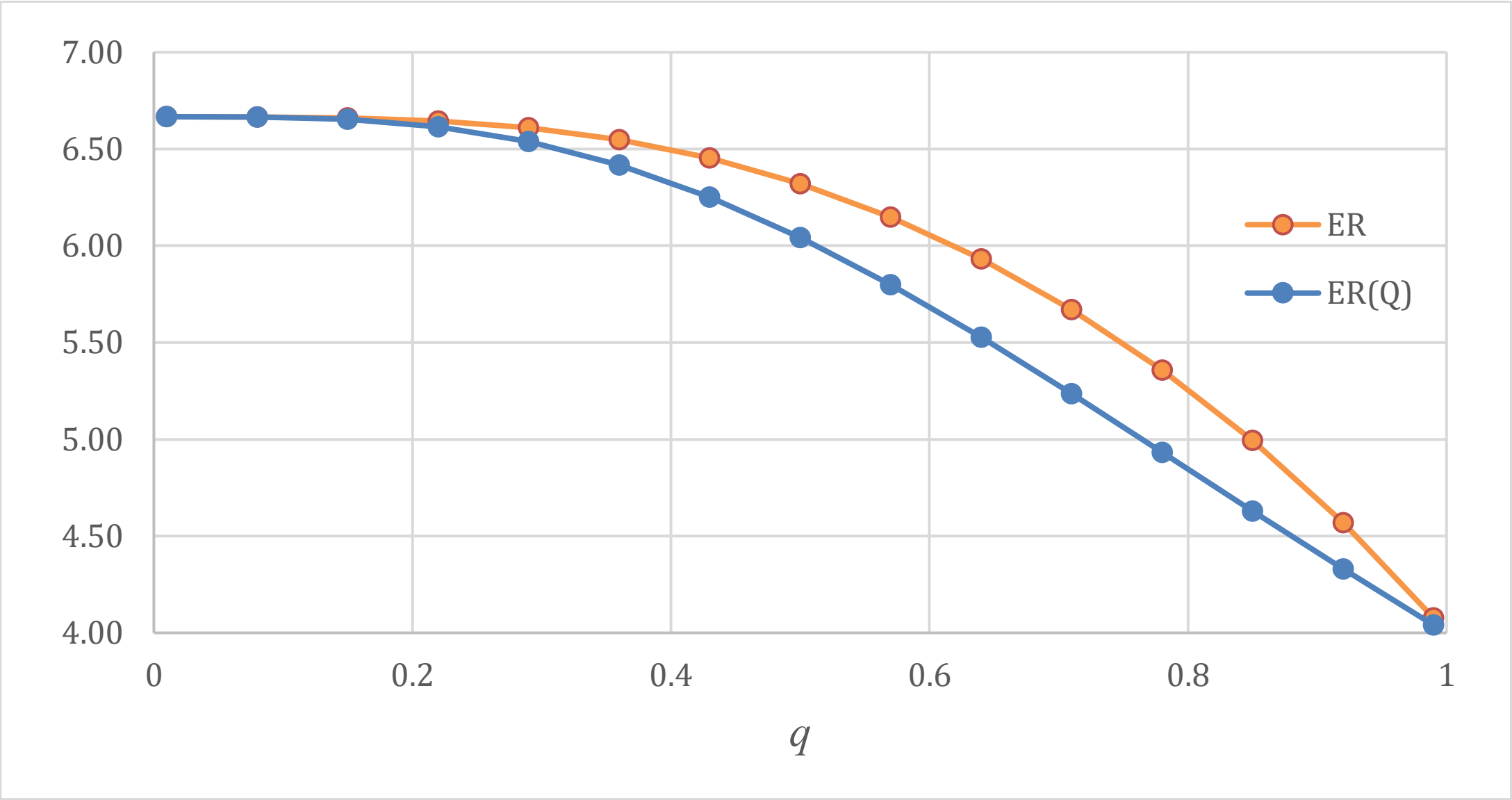

**Figure 6.** $\boldsymbol{ER}$**, and** $\boldsymbol{ER(Q)}$ **for the sequential model, under the 5-4 rule as a function of** $\boldsymbol{q}$**, given** $\boldsymbol{p = 0.75}$

## 4. **Conclusions**

We showed that the team that wins the coin toss after a tied soccer game (*A*) has a significant advantage over the team that kicks second (*B*). (Not surprisingly, *A* usually chooses to kick first). We proposed the $(m, n)$ rule to handicap *A* and thereby countervail this advantage. More specifically, when $(m, n) = (5, 4)$, *B* wins when *B* succeeds on 4 penalty kicks before *A* succeeds on 5; if both sides reach (5, 4) in the same round, the contest is decided by sudden-death rounds. There is no need for sudden-death rounds with the sequential model—when the teams alternate shooting, and the first team to reach 5 (*A*) or 4 (*B*) goals wins—but it yields almost the same results as using rounds for the probability of winning and the expected length of shootouts, as practiced today.

When *A*'s probability of scoring is 0.75 and *B*'s is 0.60, (5, 4) equalizes the probability of each side's winning the penalty shootout. But different values of $(m, n)$ may put the teams more on a par—this needs to be tested—which may include making $m > 5$ or $m - n > 1$. Although handicapping a player or team that is advantaged has proved to be a good way of creating balance in other sports (e.g., golf), another approach to making penalty shootouts more even-handed, and soccer fairer, is to vary the order of kicking in the shootout, applying, in particular, the so-called catch-up rule.[4]

---

[4] The catch-up rule prescribes that a team that is unsuccessful on a round, when the other team is successful, kick first on the next round, giving it a better chance of catching up (Brams and Ismail, 2018; Brams et al., 2018; Csató, 2021; for other proposals to revise the order of kicking in penalty shootouts, see, e.g., Echenique, 2017, Rudi et al., 2020, Anbarci et al., 2021, and Lambers and Spieksma, 2021). We think that the $(m, n)$ rule may be more appealing as a reform than the catch-up rule, because it does not tamper with the order of kicking, which may be confusing for fans to keep track of, but only the targets that each team must reach to win.